\documentclass[fleqn,usenatbib]{mnras}
\usepackage{newtxtext,newtxmath}
\usepackage[T1]{fontenc}

\usepackage{graphicx}	
\usepackage{amsmath}	
\usepackage{orcidlink}
\usepackage{multirow}

\newcommand{\cmfast}{\texttt{21cmFAST}}
\newcommand{\cmmc}{\texttt{21CMMC}}
\newcommand{\zreion}{\texttt{zreion}}
\newcommand{\nf}{x_{\text{HI}}}

\newcommand{\cv}{\texttt{CV}}
\newcommand{\fsp}{\texttt{FS}}
\newcommand{\miz}{\texttt{MI}}
\newcommand{\zr}{\texttt{ZR}}

\title[]{Mitigating Simulator Dependence in AI Parameter Inference for the Epoch of Reionization: The Importance of Simulation Diversity}
\author[]{Jasper Solt$^1$ \orcidlink{0009-0005-5818-7423}, Jonathan C. Pober$^1$ \orcidlink{0000-0002-3492-0433}, Stephen H. Bach$^2$ \orcidlink{0000-0003-3857-3560}
\\
$^1$Department of Physics, Brown University, Providence RI, USA\\
$^2$Department of Computer Science, Brown University, Providence RI, USA
}

\begin{document}
\label{firstpage}
\pagerange{\pageref{firstpage}--\pageref{lastpage}}
\maketitle

\begin{abstract}
The 21cm signal of neutral hydrogen contains a wealth of information about the poorly constrained era of cosmological history, the Epoch of Reionization (EoR). Recently, AI models trained on EoR simulations have gained significant attention as a powerful and flexible option for inferring parameters from 21cm observations. However, previous works show that AI models trained on data from one simulator fail to generalize to data from another, raising doubts about AI models' ability to accurately infer parameters from observation. We develop a new strategy for training AI models on cosmological simulations based on the principle that increasing the diversity of the training dataset improves model robustness by averaging out spurious and contradictory information. We train AI models on data from different combinations of four simulators, then compare the models' performance when predicting on data from held-out simulators acting as proxies for the real universe. We find that models trained on data from multiple simulators perform better on data from a held-out simulator than models trained on data from a single simulator, indicating that increasing the diversity of the training dataset improves a model's ability to generalize. This result suggests that future EoR parameter inference methods can mitigate simulator-specific bias by incorporating multiple simulation approaches into their analyses. 
\end{abstract}

\begin{keywords}
cosmology: dark ages -- cosmology: reionization -- cosmology: first stars -- software: machine learning
\end{keywords}

\section{Introduction}
The Epoch of Reionization (EoR) is a period of time between roughly $z=15$ and $z=6$ during which the Intergalactic Medium (IGM) underwent a phase change from fully neutral to fully ionized, driven by the radiation from the first galaxies following Cosmic Dawn. The 21cm signal emitted by neutral hydrogen is expected to yield rich information about this era of cosmological history, as it serves as a tracer for the distribution of neutral hydrogen in the universe and can therefore directly map the fraction of neutral hydrogen $\nf$ over time. Yet the EoR remains poorly constrained, even when compared to more far-flung time periods such as recombination ($z\approx 1100$). This is largely due to the difficulty of observing the 21cm signal: bright foregrounds, low signal-to-noise, and complex instrumental effects make untangling the signal from the data challenging. 

Ongoing and future low-frequency radio telescope experiments, such as the Low Frequency Array (LOFAR, \citealt{Haarlem2013}), the Murchison Widefield Array (MWA, \citealt{Tingay2013}), the Hydrogen Epoch of Reionization Array (HERA, \citealt{DeBoer2017}), and the Square Kilometer Array (SKA, \citealt{Braun2015}, \citealt{Koopmans2015}), promise to improve signal constraints and eventually provide a direct detection of the EoR. However, actually extracting parameter constraints from observations remains a difficult task. 

Existing analytic models fail to capture all the complexity of the physics involved in reionization, and are therefore not preferred for extracting parameter values from observation. Therefore, we use simulations to constrain EoR data: given a physical model and set of assumptions, we search the input space for a set of parameters that best fit our observations. However, to what degree our simulations generalize to the real universe is an open question. This question forms the core motivation for this work.

The current state-of-the-art method for 21cm parameter constraint is \cmmc, a Markov Chain Monte Carlo tool designed to search the parameter space of \cmfast, its corresponding model of reionization (\citealt{Greig2015}). This approach has the advantage of being based on well-understood Bayesian statistics and therefore inherently including uncertainty information in the analysis. However, \cmmc\ has its drawbacks. \cmmc\ is limited to searching the input space of \cmfast, the parameters of which are approximations to more complicated physics and therefore require significant interpretation when applied to data from other simulators. Additionally, \cmmc\  uses the power spectrum as a summary statistic, which does not contain all information about the ionization field due to non-Gaussianities in the signal and therefore may be sacrificing valuable information (\citealt{Majumdar2018}, \citealt{Shimabukuro2016}). 

As a result, training AI models on simulated data has gained significant recent attention as an option for 21cm parameter analysis due to the ability of AI models to regress on complex, noisy data. In contrast to \cmmc, AI models are flexible in that they can efficiently regress on many data formats, such as the power spectrum as well as raw image data. AI models also have the potential to synthesize information from many different sources of data, removing the inherent dependency on any one model of reionization.

AI models can only be as good as the data they are trained on; therefore, it is worth taking a closer look at the EoR simulators that underpin our inference methods. Simulators of the EoR can be divided into two loose categories: ``numeric'' and ``semi-numeric''. Broadly speaking, numeric simulators make fewer assumptions and take into account more complex physics (such as radiative transfer and hydrodynamic effects) than semi-numeric simulators. In contrast, semi-numeric simulators make smart assumptions to reduce computational cost, allowing more data at larger simulation volumes to be produced (a review of different reionization models can be found in \citealt{Gnedin2022}). Because parameter inference requires large amounts of high-volume data (simulation box lengths must be $\geq300$ Mpc for large-scale modes to converge; see \citealt{Iliev2014}, \citealt{Kaur2020}), semi-numeric simulators are preferred for this type of analysis. 

All semi-numeric simulators of reionization will have some inherent non-negligible error, due to the assumptions and approximations characteristic of each individual simulator. Because of this, semi-numeric simulators are not perfectly reflective of the physics of the real universe. This raises a concern: can parameter inference methods based on imperfect simulations generalize to real data? 

This concern over generalizability is not unfounded: \cite{Berklas2025} finds that \cmmc, the current state-of-the-art method for EoR parameter inference, fails to accurately infer the parameters of different versions of \cmfast, while \cite{Zhou2022} find that AI models trained on one semi-numeric simulator fail to generalize to another. A key question, then, is this: how do we ensure AI models trained on imperfect simulations can make accurate predictions about real-life 21cm observational data? 

A classic strategy in AI research involves increasing the diversity of the training data to avoid overfitting to spurious features. By increasing variation within the training data, incorrect or irrelevant information is ``averaged-out'', leaving only useful commonalities behind. Because all semi-numeric simulators make different approximations, each is flawed in unique ways. Incorporating multiple simulators into our analysis, therefore, may reduce simulator-specific bias and improve accuracy when predicting on data from unseen distributions, such as real-world observation data. This work investigates that claim.

This paper is structured as follows. Section \ref{sec:method} outlines our approach to testing AI model generalizability, the generation parameters for our four datasets, and the structure of our model. Section \ref{sec:results} compares the parameter inference on out-of-distribution data for models trained on one, two, and three simulators, and finds that on average generalizability improves with added diversity. Finally, Section \ref{sec:discussion} discusses these results along with their implications for the broader field of EoR parameter inference.

\subsection{Related Work}
\label{sec:related_work}
The simulator-dependence of different EoR parameter inference methods remains largely uninvestigated. \cite{Berklas2025} finds that the state-of-the-art EoR parameter inference method \cmmc\ fails to recover parameters from different variations of the same semi-numeric simulator, thus failing to generalize to an unseen data distribution. \cite{Zhou2022} tests the generalizability of AI models given two datasets generated from different simulators of reionization. They find that AI models trained on one simulator typically fail to generalize to another. They also find that training an AI model on data from both simulators results in good performance on each. In this work, we will investigate the generalizability across four different datasets, allowing us to pursue a fundamentally new kind of analysis, answering the questions ``does training an AI model on data from two reionization simulators improve performance on data from a third reionization simulator?'' and ``is there a trend between the number of simulators represented in the training data and the AI model's ability to generalize to data from an unseen simulator?'' 

More broadly speaking, many works have investigated AI's capacity for image-based inference of EoR parameters. These largely fall into one of two camps: experiments in which the initial density fields are varied, but the predicted parameters are simulator-specific and therefore do not map easily to data generated from other simulators (\citealt{Gillet2019}, \citealt{Kwon2020}, \citealt{Neutsch2022}, \citealt{Zhao2022}, \citealt{Prelogovic2021}, \citealt{Hassan2020}), or experiments where the predicted parameters are model independent, but the data is generated from a single initial density field (\citealt{LaPlante2019}, \citealt{Billings2021}, \citealt{Zhou2022}). The former makes it difficult to determine a model's ability to generalize across distributions (for example, if we trained an AI model to predict the value of \cmfast's $\zeta$ parameter, it would not be possible to evaluate its performance on \zreion\ data, which has no such parameter) while the latter can cause the model to overfit to specific regions or spatial features of the data (\citealt{Sooknunan2024}). This work uses unique density fields for every reionization history generated while also predicting on model-agnostic parameters.

In addition to using AI to analyze EoR image data, there has been significant research into AI power-spectrum based parameter estimation (\citealt{Doussot2019}, \citealt{Shimabukuro2017}, \citealt{Choudhury2022}, \citealt{Tiwari2022}, \citealt{Shimabukuro2022}). This would essentially serve as an alternative to \cmmc, being more computationally efficient and possibly more accurate while sacrificing the uncertainty information inherent to the Bayesian framework. 

Finally, AI is being integrated into EoR data analysis pipelines. In seeking to reduce the bias of LOFAR's Gaussian Process Regression (GPR) method of signal separation, \cite{Mertens2023} replaces their covariance prior model with a learned autoencoder, coining the term ``ML-GPR'' for their method. This pipeline has already been incorporated in analysis of LOFAR data (\citealt{Mertens2025}, \citealt{Munshi2025}). 

\section{Method} 
\label{sec:method}

In recent years, AI research has trended towards a data-centric (as opposed to model-centric) approach, emphasizing the curation of high-quality training data over the development of better model designs. This shift comes as a result of a growing body of research showing that the architecture of the model matters much less than the quantity and quality of the data (\citealt{Zha2025}). Following this approach, this work therefore investigates how the data used for parameter inference affects the accuracy of the inferred parameters.

Increasing training data diversity allows AI models to focus on commonalities while discarding spurious and conflicting features of individual simulators, thereby creating a more informed and cohesive picture of the distribution. Therefore, we propose that the greater the number of EoR simulation algorithms included during analysis, the more universal properties of the set can be inferred by our parameter inference method of choice. This concept of increasing dataset diversity to improve generalizability motivates our approach. 

To study how diversity affects out-of-distribution performance, we train models on data from one, two, and three simulators of reionization before predicting on data from a held-out simulator. In this work, we use the term ``out-of-distribution'' to refer to data generated from any simulator not represented in the training data. Following this, we compare the performance of the single, double, and triple simulator models to see if the out-of-distribution performance has improved. If generalizability to an unknown distribution does improve with increased data diversity, that implies AI parameter inference of real-world EoR data would benefit from a large and varied suite of training data, drawn from many different simulators of reionization.

Section \ref{sec:data} describes our simulation suite and dataset design, while Section \ref{sec:model} briefly covers the structure of our AI model.

\begin{figure}
    \includegraphics[width=\columnwidth]{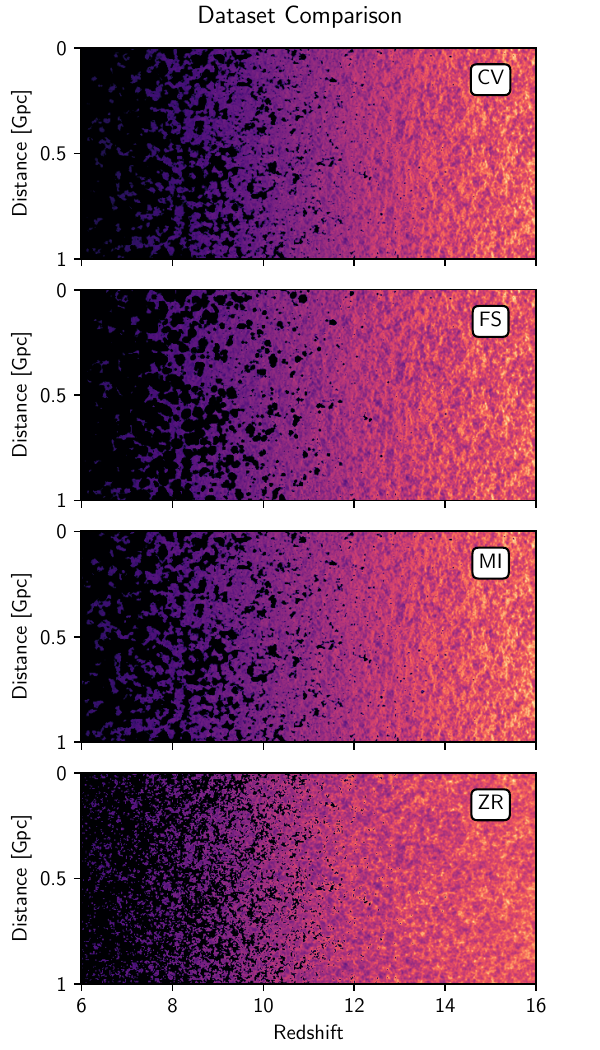}
    \caption{A comparison of the different algorithms used in this work to simulate the evolution of the EoR brightness temperature field. For this comparison, each EoR history uses the same underlying density field, though in the actual datasets no two instances share the same initial density field. For this figure, input parameters were adjusted in order to produce instances with similar reionization histories while still showcasing the qualitative differences between simulators. For example, note how Dataset \cv\ results in smaller more jagged bubbles while Dataset \fsp\ produces comparatively larger and rounder bubbles, reflecting their respective bubble-flagging methods.}
    \label{fig:compare_datasets}
\end{figure}
        
\begin{table*}
 \label{tab:compare_datasets}
 \begin{tabular}{lccccc}
  \hline
  Dataset Name & Abbrev. & Source Simulator & Mass-dependent $\zeta$ & Bubble Finding Algorithm & Box Length (Mpc) \\
  \hline
  ``Central-Voxel'' & \cv\ & \cmfast & Yes & Central-voxel flagging & 1000  \\ 
  ``Full-Sphere'' & \fsp\ & \cmfast & Yes & Full-sphere flagging & 1000  \\ 
  ``Mass-Independent Zeta'' & \miz\ & \cmfast & No & Central-voxel flagging & 2000  \\ 
  ``Zreion'' & \zr\ & \zreion & - & - & 1000 \\ 
  \hline
 \end{tabular}
 \caption{A summary of the differences between Datasets \cv, \fsp, \miz, and \zr. A more detailed list of generation parameters can be found in Appendix \ref{appendix:dataset_params}.}
\end{table*}

\subsection{Data}
\label{sec:data}
In order to test how a model performs on an unseen distribution, four sets of data were generated, each of which used different approximations to calculate the evolving $\nf$ field. Two different EoR simulation packages were used to generate data: \cmfast\ and \zreion. \cmfast\ was used to generate Datasets \cv\ (``Central-Voxel''), \fsp\ (``Full-Sphere''), and \miz\ (``Mass-Independent Zeta''). This naming scheme is based on the different versions of \cmfast\ used to generate each: Datasets \cv\ and \fsp\ use the mass-dependent $\zeta$ parameterization, but use the central-voxel and full-sphere bubble flagging methods, respectively, while Dataset \miz\ uses the mass-independent $\zeta$ parameterization\footnote{Like Dataset \cv, Dataset \miz\ also uses central-voxel bubble flagging, but this was omitted from its name to avoid a cumbersomely long abbreviation.}; more details about the differences between these versions of \cmfast\ can be found in Section \ref{sec:21cmfast}. Dataset \zr\ (``Zreion''), meanwhile, was generated using the more approximate semi-numerical simulator \zreion. 

Sections \ref{sec:21cmfast} and \ref{sec:zreion} give an overview of \cmfast\ and \zreion, the simulators used to generate our four datasets. Section \ref{sec:labels} then discusses our chosen simulator-dependent inference parameter, the duration of reionization $\Delta z$. Finally, Section \ref{sec:data_gen} outlines our data generation parameters and preprocessing steps.

\subsubsection{21cmFAST}\label{sec:21cmfast}
\cmfast\ (\citealt{Mesinger2011}, \citealt{Greig2015}, \citealt{Murray2020}) is a semi-numerical method for simulating the ionization of the IGM during the EoR. It uses Lagrangian perturbation theory to evolve the linear density field over time, then estimates the ionization field directly using an approach based on the excursion set formalism which compares the number of ionizing photons to the number of baryons over a spherical region of decreasing radius $R$. At a given redshift $z$, a region with radius $R$ centered at position $\textbf{x}$ is considered ionized when the product of the ionization efficiency $\zeta$ and the collapse fraction $f_\text{col}$ is greater than $1$:
\begin{equation}
    \label{eq:ionization_condition}
    \zeta f_\text{col} (\textbf{x}, z, R, \bar{M}_{\text{min}}) \geq 1
\end{equation}
Here, $f_\text{col}$ is the fraction of collapsed matter residing in halos larger than $\bar{M}_{\text{min}}$.

$\zeta$ can be further expressed as a product of the fraction of baryons incorporated in stars $f_*$, the escape fraction $f_{\text{esc}}$, and the number of ionizing photons per baryon in stars $N_{\gamma / \text{b}}$:
\begin{equation}
    \label{eq:zeta}
    \zeta=f_*f_{\text{esc}}N_{\gamma / \text{b}}
\end{equation}
Partial ionization is simulated by setting the ionization fraction to $\zeta f_\text{col} (\textbf{x}, z, R, \bar{M}_{\text{min}})$ when $R$ equals the cell size.

A few variations of the \cmfast\ algorithm exist; these unique formulations are used to differentiate Datasets \cv, \fsp, and \miz. One modification that can be made to the default \cmfast\ algorithm is the way in which ionized bubbles are flagged. The condition presented in Equation \ref{eq:ionization_condition} can be applied to identify ionized bubbles using two different algorithms: central-voxel flagging and full-sphere flagging. Central-voxel flagging is the default behavior, in which only the voxel at point $(\textbf{x}, z)$ is considered ionized when the condition in Equation \ref{eq:ionization_condition} is met. Full-sphere flagging, in contrast, marks every point within $R$ as ionized. There is conflicting evidence as to which flagging algorithm is more reflective of real-world physics (\citealt{Mesinger2011}, \citealt{Hutter2018}); for the purposes of this work, which algorithm is more correct is less important than the fact that both are plausible. In this work, Datasets \cv\  and \miz\ use the central-voxel bubble finding algorithm, while Dataset \fsp\ (``Full-Sphere'') uses the full-sphere algorithm (see Table \ref{tab:compare_datasets} for details).

Another modification one can make to the default \cmfast\ algorithm is whether $\zeta$ depends on the halo mass $M_\text{h}$. In the mass-independent formulation given in Equation \ref{eq:ionization_condition}, $\zeta$ is assumed to be constant, independent of $M_\text{h}$. In a new parameterization introduced by \cite{Park2019}, $f_*$ and $f_{\text{esc}}$ are redefined as functions of $M_\text{h}$, allowing $\zeta$ to vary with the halo mass. This mass-dependent $\zeta$ parameterization more accurately reflects the effects of the mass of a host halo on star formation rate and the emission of ionizing photons. This work uses this parameterization for Datasets \cv\ and \fsp, while using the mass-independent $\zeta$ formulation for Dataset \miz\ (``Mass-Independent Zeta''). Therefore, while these three datasets are all generated from \cmfast, they each use unique parameterizations of \cmfast, and therefore represent distinct distributions.

\subsubsection{zreion}
\label{sec:zreion}
\zreion\footnote{\url{https://github.com/plaplant/zreion}} is an improved implementation of Reionization at Large Scales (RLS), a method presented in \cite{Battaglia2013} developed as a lightweight way to simulate reionization for very large box sizes. \zreion\ assumes the redshift of reionization field $\delta_z$ is a linear tracer of the matter overdensity field $\delta_m$ at large scales. The model is defined by the bias parameter $b^2_{zm}(k)$ which relates the two fields in Fourier space:
\begin{equation}
    b^2_{zm}(k) = \frac{P_{zz}(k)}{P_{mm}(k)} = \frac{b_0^2}{\left(1+\frac{k}{k_0}\right)^{2\alpha}}
\end{equation}
where $P_{zz}$ and $P_{mm}$ are the power spectra of $\delta_z$ and $\delta_m$ respectively. Setting $b_0 = \frac{1}{\delta_c} = 0.593$ ($\delta_c$ being the critical overdensity in the spherical collapse model) leaves \zreion\ three free parameters with which to model reionization: the wavenumber scaling factor $k_0$, the power law index $\alpha$, and the mean redshift of reionization $\bar{z}$. Unlike \cmfast, \zreion\ does not simulate partial cell ionization, instead expressing the neutral fraction $\nf$ as a binary 0 or 1.

As \zreion\ does not have a built-in way to simulate the evolution of the linear density field, all density fields were generated using \cmfast. 

Because \zreion\ treats the ionization field as a linear function of the density field, it fails to capture aspects of reionization other simulators can, such as the characteristic growth of ionized bubbles around individual radiation sources. However, these differences become less important on large scales (\citealt{Battaglia2013}).

While \zreion\ is less physically motivated in its parameterization than \cmfast, its usefulness to this work lies in the fact that it represents a significantly different set of possible reionization histories than those which can be generated with \cmfast. This allows us to determine the extent of our model's ability to generalize to new distributions.

\subsubsection{Labels}
\label{sec:labels}
\begin{figure}
    \includegraphics[width=\columnwidth]{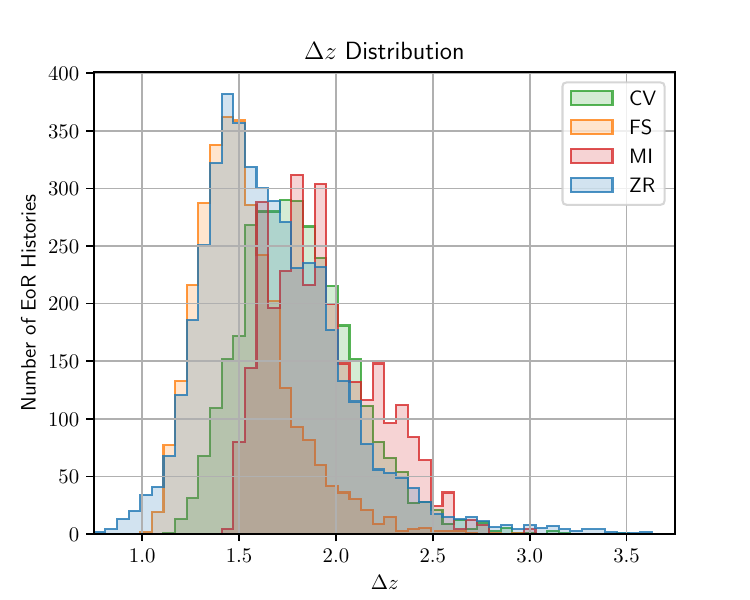}
    \caption{Distribution of $\Delta z$ across all four datasets. It is important that these distributions overlap, as AI models traditionally struggle to predict on label ranges outside that of their training data (\protect\citealt{Sooknunan2024}).}
    \label{fig:dur_distribution}
\end{figure}

In this work, we choose to predict on the duration of reionization $\Delta z = z_{25}-z_{75}$, where $z_{25}$ and $z_{75}$ are the redshifts at which 25\% and 75\% of the neutral hydrogen is ionized respectively. This choice was motivated by the need for a prediction parameter with a concrete physical interpretation. Since no analytic model fully describes reionization, any set of abstract input parameters will inherently be nonphysical. In contrast, $\Delta z$ is a value with real physical meaning, which can be determined from the neutral fraction history of any reionization simulation. In addition, $\Delta z$ also has the advantage of being simulator independent (see discussion of simulator dependent vs. simulator independent parameters in Section \ref{sec:related_work}). How $\Delta z$ relates to other reionization parameters, such as the Cosmic Microwave Background optical depth $\tau$, is a question of ongoing research. Nevertheless, $\Delta z$ remains an important quantity in and of itself, especially given the limited existing observational constraints on the EoR. 

For Datasets \cv, \fsp, and \miz, $\Delta z$ was calculated from the global neutral fraction, a value automatically tracked by \cmfast. For Dataset \zr, the neutral fraction history was calculated directly from the ionization field. 

Figure \ref{fig:dur_distribution} shows a histogram of the $\Delta z$ values for all four datasets. During data generation, the range of allowed input parameters was adjusted to ensure the $\Delta z$ distributions overlapped (see Appendix \ref{appendix:dataset_params} for a detailed breakdown of the range of dataset generation parameters). Overlapping output ranges between the training and testing datasets has been found to play an important role in AI model performance (\citealt{Sooknunan2024}). While the distributions in Figure \ref{fig:dur_distribution} are not identical, there exists a region where all four datasets overlap such that our models are rarely predicting a $\Delta z$ value outside the range seen during training. 

\subsubsection{Data Generation and Preprocessing}
\label{sec:data_gen}

Histories were generated for Datasets \cv, \fsp, and \miz\ using the semi-numerical simulation package \cmfast, with parameters randomly sampled over the ranges listed in Appendix \ref{appendix:dataset_params}. 

Initially, 4399 histories were generated for Dataset \cv, 3981 histories were generated for Dataset \fsp, and 1031 histories were generated for Dataset \miz. Datasets \cv\ and \fsp\ were generated using a simulation box length of $1000$ Mpc and a plane-of-sky resolution of $256\times256$, resulting in a cell size of $3.9$ Mpc. Dataset \miz\ was generated using a simulation box length of $2000$ Mpc and a plane-of-sky resolution of $512\times512$, resulting in the same cell size of $3.9$ Mpc. Reionization histories for all three datasets were evolved over the redshift range $[6.0, 16.0]$. 

We used \cmfast's inbuilt interpolator to knit our evolving brightness temperature fields into brightness temperature lightcones. Here, the term ``lightcone'' refers to a simulated reionization scenario in which the xy-plane represents the plane of the sky, and the z-axis represents evolving redshift. These lightcones were then trimmed to span the redshift range [$6.0$, $14.7$], resulting in a z-axis length of 512.\footnote{$14.7$ was chosen as the maximum redshift simply because it corresponds to the redshift slice at index 512 along the z-axis, and is otherwise completely arbitrary.}

To ensure each lightcone had a defined value for $\Delta z$, any generated histories where $z_{25}\geq16.0$ or $z_{75}\leq6.0$ were discarded at the time of generation. This resulted in the final dataset sizes of 3445 lightcones for Dataset \cv, 3062 lightcones for Dataset \fsp, and 740 lightcones for Dataset \miz. This corresponds to a roughly $25\%$ loss in data.

The discrepancy in simulation volume between Datasets \miz\ and Datasets \cv\ and \fsp\ is due to the fact that Dataset \miz\ was generated first; the box length was halved for subsequent datasets because the simulation volume was determined to require too much computational time to be feasible for all four datasets. Because the cell size was consistently $3.9$ Mpc across all datasets, Dataset \miz\ lightcones are essentially twice as large as those from other datasets along the x and y axes. Therefore, to make all data cubes have the same shape, Dataset \miz\ lightcones were quartered on the xy-plane, resulting in a final dataset size of 2960 reionization histories for Dataset \miz. While it is possible the increased simulation volume impacted the resulting brightness temperature field due to the availability of more spatial modes, we do not consider this to be a problem, as this would only serve to diversify our four datasets further. 

Dataset \zr\ is based on a different semi-numerical simulation package than the other datasets, and was therefore generated using a different process. Because \zreion\ does not include a way to generate evolving cosmological density fields, the density fields for Dataset \zr\ were created and perturbed over time using \cmfast. 4538 perturbed density fields were generated in this way, using a simulation box length of $1000$ Mpc and a plane-of-sky resolution of $256\times256$, resulting in a cell size of $3.9$ Mpc, identical to that of our three other datasets. These perturbed density fields were then interpolated to form density lightcones spanning the redshift range [$6.0$, $14.7$], identically to how lightcones were generated for the previous three datasets. These density lightcones were then used as inputs to the \zreion\ simulator, with parameters randomly sampled over the ranges and priors listed in Appendix \ref{appendix:dataset_params}. Unique density fields were generated for each reionization history across all datasets (see Section \ref{sec:related_work} for a discussion on the importance of unique density fields).  

Building on previously published approaches (\citealt{LaPlante2019}, \citealt{Zhou2022}), each lightcone was divided into 30 $256\times256$ mean-subtracted slices of the brightness temperature field, sampled evenly along the full redshift range of [$6.0$, $14.7$]. We refer to these redshift-sliced lightcones as ``data cubes.''

Besides mean subtraction, no foregrounds, sources of noise, or instrumental effects were modeled in any of the data. Because of this, one might naively believe that $\Delta z$ can be trivially calculated by assuming all cells where the differential brightness temperature field $\delta T_\text{b}(\mathbf{x}, z) = 0$ are ionized, giving us an estimate of $\nf(z)$. This is a decent assumption for \zreion, but it does not hold for \cmfast\ data due to the fact that \cmfast\ simulates partial cell ionization as described in Section \ref{sec:21cmfast}. In Section \ref{sec:results}, we will show that our model struggles to generalize to data from unseen simulators even without foregrounds or noise modeled in the data. Therefore, we treat this work as an exploratory proof-of-concept for a method to reduce simulator specific bias in AI models by increasing training data diversity, leaving explorations of how sources of information loss affect model performance on out-of-distribution data for future works.

Prior to training, data cubes were normalized to the interval $[0,1]$ for ease of regression. Each dataset was divided into training, validation, and testing subsets, with 2400, 280, and 280 data cubes respectively. This size was chosen because our smallest dataset, Dataset \miz, includes $2400+280+280=2960$ data cubes; in this way, we control for training dataset size. 

Because every reionization history was initialized with a unique random seed, there is no cross-contamination of density field information across different splits of the data. This is essential to prevent the AI model from learning to look for specific regions of structure in the data (\citealt{Sooknunan2024}). This is even the case with the subdivided Dataset \miz\ lightcones, as the four quadrants of the xy plane across which each lightcone was split were not correlated in any way.

For models trained on multiple simulators, the size of the training dataset was limited to 2400 cubes by splitting the pool of training data proportionally between the represented simulators, i.e., models trained on data from two simulators had training datasets composed of 1200 data cubes from each, and models trained on data from three simulators had training datasets composed of 800 data cubes from each. This choice was made to control for the size of the training dataset, which is known to impact model performance. This dataset splitting was not done for the test and validation sets, which were kept at 280 data cubes per simulator. This was done so that the performance of each model on the test data from each simulator could be equivalently compared. 

\subsubsection{k-Fold Validation} 
k-fold cross-validation was performed over 8 folds to ensure the results are not dependent on different train / validation / test partitions of the dataset, via the following procedure:

\begin{enumerate}
    \item Each simulator dataset of size 2960 were split into 10 groups of size 280, leaving 160 data cubes ungrouped. These ungrouped data cubes were considered ``remainders'' and were never folded into the test / validation groups. This was done to preserve the train / validation / test split of 2400 / 280 / 280; having a training dataset of a length divisible by 2 and 3 is necessary to include equal proportions of data from each simulator in the case of models trained on data from two and three simulators.
    \item For fold $i$ of 10, group $i$ was selected as the test group and group $(i+1)\mod{10}$ was selected as the validation group. This leaves all other groups (plus the 160 ``remainders'') available to be used in the training data. For single-simulator models, this data was used as the training group.
    \item In the case of models trained on data from two or three simulators, where the training dataset is already split into groups of 1200 and 800 data cubes from each simulator respectively, we also fold along this partition, taking data cubes from each simulator dataset starting at the index for group $(i+2)\mod{10}$. In this way, the training data changes with every fold, even when the test and validation groups do not require it to change (aka, when the test and validation groups are in the proportion of the simulator dataset that would not have been used for training anyway).
\end{enumerate}

We report our results as mean $\pm$ standard error across all 8 folds.

\subsection{Model}
\label{sec:model}
\begin{figure*}
    \includegraphics[width=\textwidth]{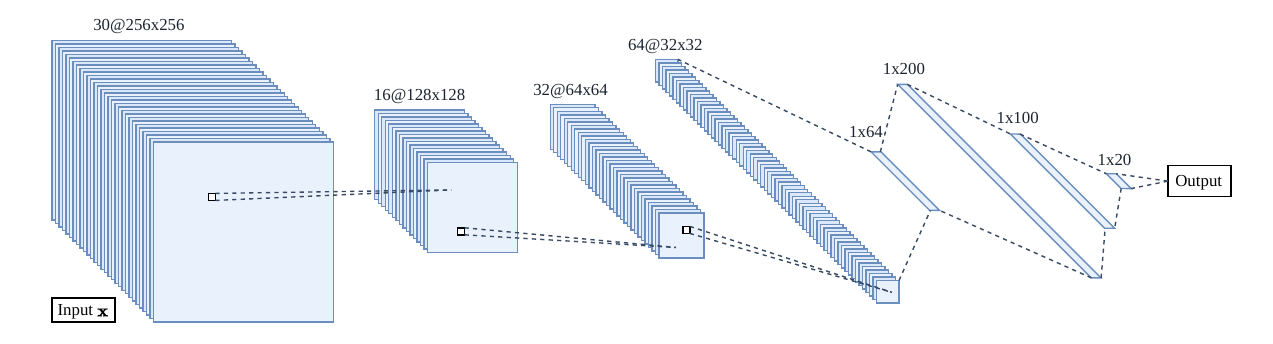}
    \caption{A simple diagram of our CNN. Our model consists of three 2D convolutional downsampling layers, a global pooling layer, and three linear layers, with a resulting scalar output. Our model takes data cubes of size $30\times256\times256$ as inputs to the network. Each cube is made up of a set of 30 brightness temperature slices sampled evenly along the redshift axis, treated as color channels such that the convolutional kernel convolves over the xy plane.}
    \label{fig:model_architecture}
\end{figure*}

\begin{table}
    \centering
    \begin{tabular}{l|c}
        Parameter & Value \\
        \hline
        Batch Size & 64\\
        Stage I Learning Rate & $5\times 10^{-3}$ \\
        Stage I Epochs & 1000 \\
        Stage II Learning Rate & $1\times 10^{-3}$ \\
        Stage II Epochs & 2000 \\
    \end{tabular}
    \caption{Model learning parameters used during training.}
    \label{tab:model_learning_params}
\end{table}

Our model is a convolutional neural network (CNN) based on the architecture used in \cite{Zhou2022}, and is summarized in Figure \ref{fig:model_architecture} (a more detailed breakdown of our model's architecture can be found in Appendix \ref{appendix:model_architecture}). It consists of three downsampling 2D convolutional layers, a global pooling layer, and three linear layers. Our model takes data cubes with dimensions $30\times256\times256$ as inputs. The 30 redshift slices are stacked along the channel dimension such that the convolutional kernel convolves over the xy plane. In this way redshift information is treated as analogous to color channel information for a more traditional image-analysis CNN. In some previous research, 3D convolutional layers have shown promise in more effectively integrating information along the redshift axis (\citealt{Zhao2022}), but we do not explore that here. 

Models were trained to convergence using the learning hyperparameters summarized in Table \ref{tab:model_learning_params}. The batch size and learning rate were optimized for via a parameter grid search prior to training. Training took place in two stages, with a stepped-down learning rate after Epoch 1000.  The loss function was Mean Squared Error (MSE). The same architecture and learning parameters were used across all models trained. 

\section{Results} 
\label{sec:results}
\begin{figure*}
    \includegraphics[width=0.8\textwidth]{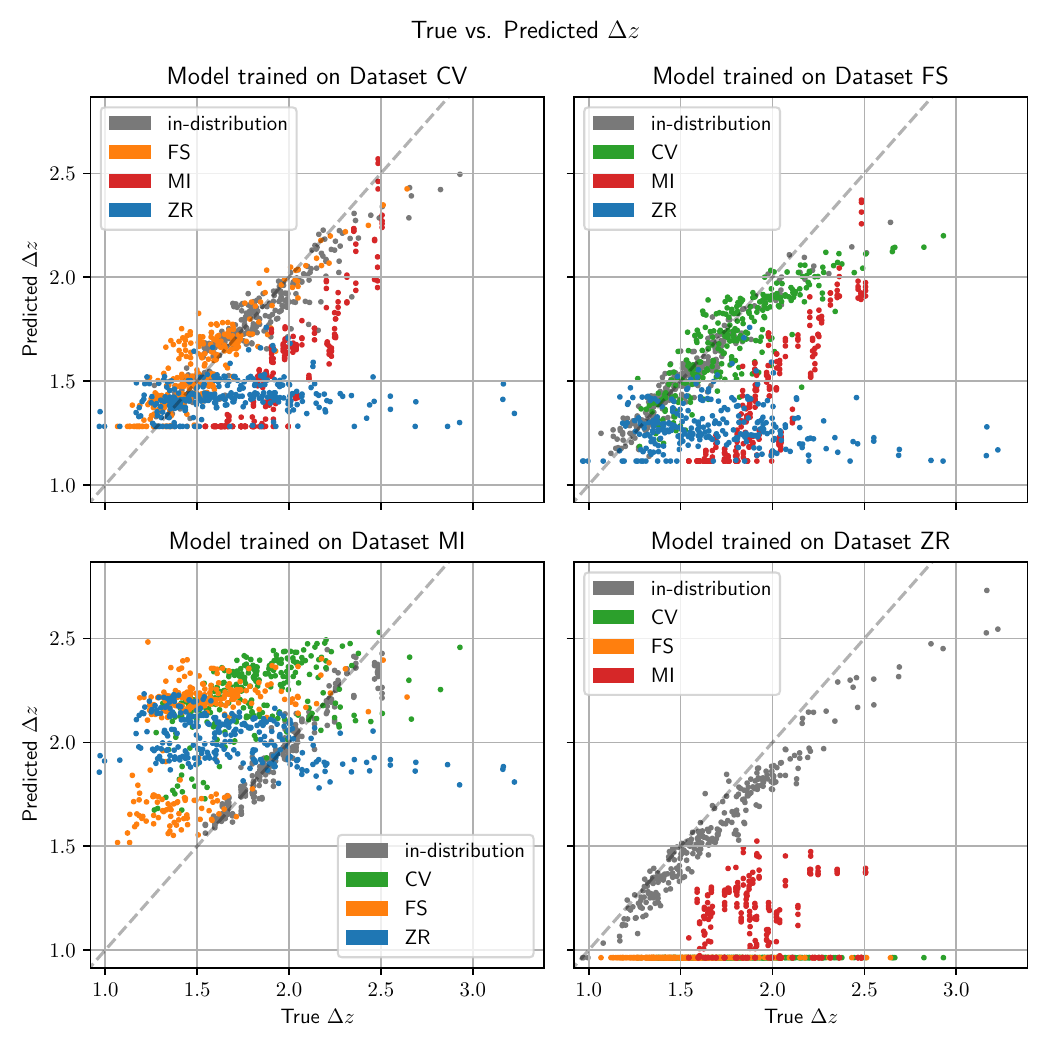}
    \caption{Predicted vs. true $\Delta z$  of single-dataset models across all four datasets. Note the clear difference between in-distribution and out-of-distribution performance. For the models trained on Datasets \cv, \fsp, and \zr, the model will not predict values of $\Delta z$ below the minimum value seen during training, resulting in a horizontal line of predictions at the minimum. This is likely due to the sparsity of the trained model's latent space, an issue that could be rectified by tweaking the model architecture or further hyperparameter optimization, though doing so is unlikely to result in a significant boost in accuracy.}
\label{fig:single_dataset_ood_scatter}
\end{figure*}
We find that, when trained on data drawn from only one distribution, our models do not generalize to data from a second unseen distribution. Figure \ref{fig:single_dataset_ood_scatter} shows the predicted vs. true $\Delta z$ of both in-distribution vs. out-of-distribution test cubes for models trained on Datasets \cv, \fsp, \miz, and \zr. Here, we choose to primarily report average MSE rather than average percent error, due to the the fact that the $\Delta z$ distributions of our datasets are all slightly different, which means percent error cannot be used as a relative measure to compare between them. The single-dataset model is able to recover the value of $\Delta z$ from in-distribution reionization histories (shown in gray) with an average MSE of $0.01$. However, when asked to predict the value of $\Delta z$ for out-of-distribution data cubes (shown in color), the model does so with substantial bias, with an average MSE of $0.32$ across all out-of-distribution dataset / model pairs. The one exception to the poor out-of-distribution performance is that models trained on Dataset \cv\ seem to perform well on Dataset \fsp\ and vice versa; a more in-depth discussion of how different simulators generalize to one another can be found in Section \ref{sec:x_sim_performance_transfer}. 

We emphasize that \textit{none of these test points were seen during training}, as the data was split into training / validation / testing subsets prior to any training, and data from the test subsets were never used to train models; the distinction between an in-distribution and out-of-distribution data cube comes from which simulator was used to generate it, not whether it specifically was included in the training data. 

This failure to generalize mirrors the findings of \cite{Zhou2022}, with more datasets and unique density fields. We demonstrate here that the added complexity and diversity introduced by unique underlying density fields is not sufficient to allow the models to interpolate to unseen distributions, thus motivating our investigation into diversifying our training data with multiple simulators.

\begin{figure*}
    \includegraphics[width=0.8\textwidth]{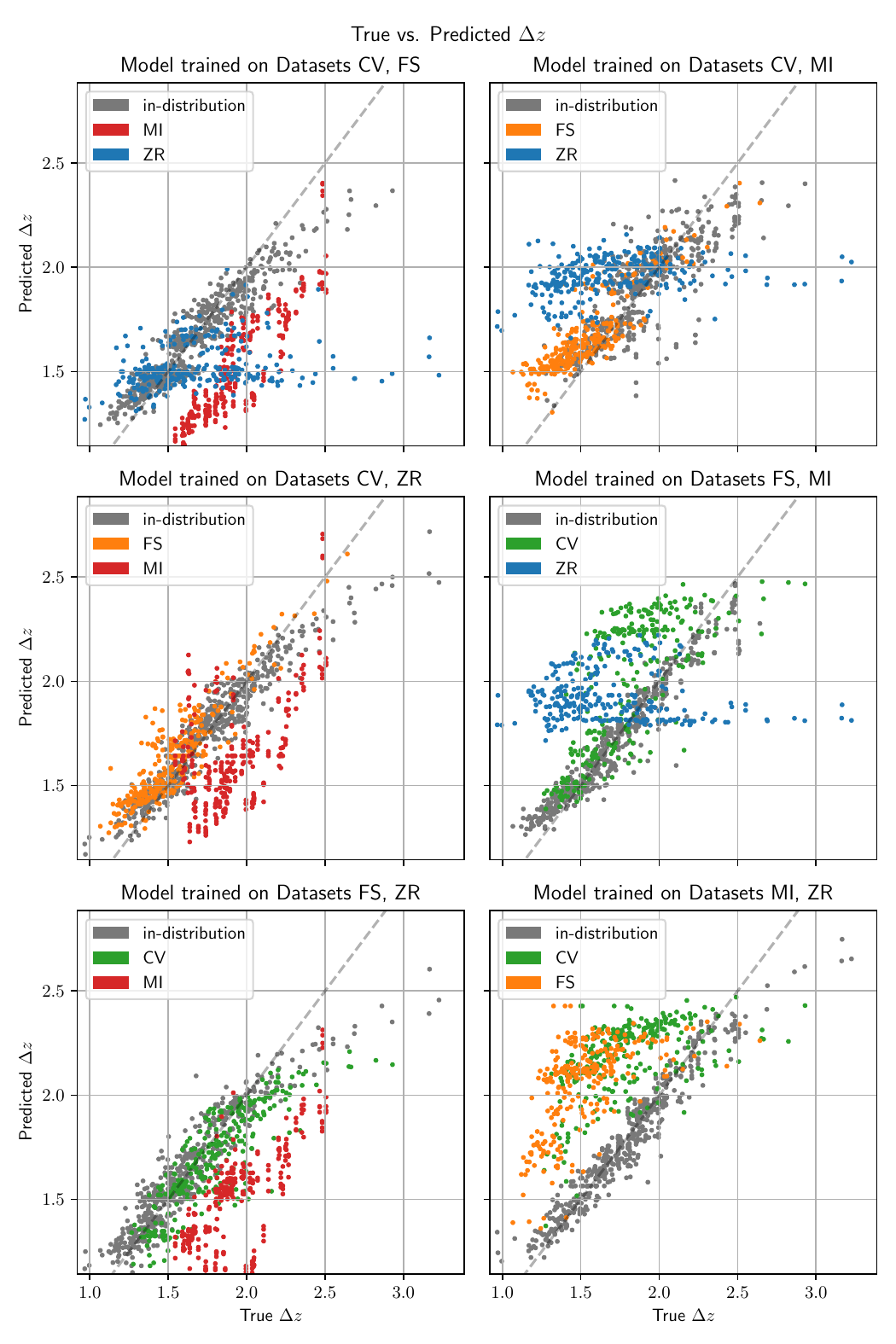}
    \caption{Predicted vs. true $\Delta z$ for models trained on two datasets. Note that unlike the training data, which was split equally between the two simulators used during training, the test datasets plotted here remain at a fixed size, in order to better compare the test performance across models.}
\label{fig:double_dataset_ood_scatter}
\end{figure*}

\begin{figure*}
\includegraphics[width=0.8\textwidth]{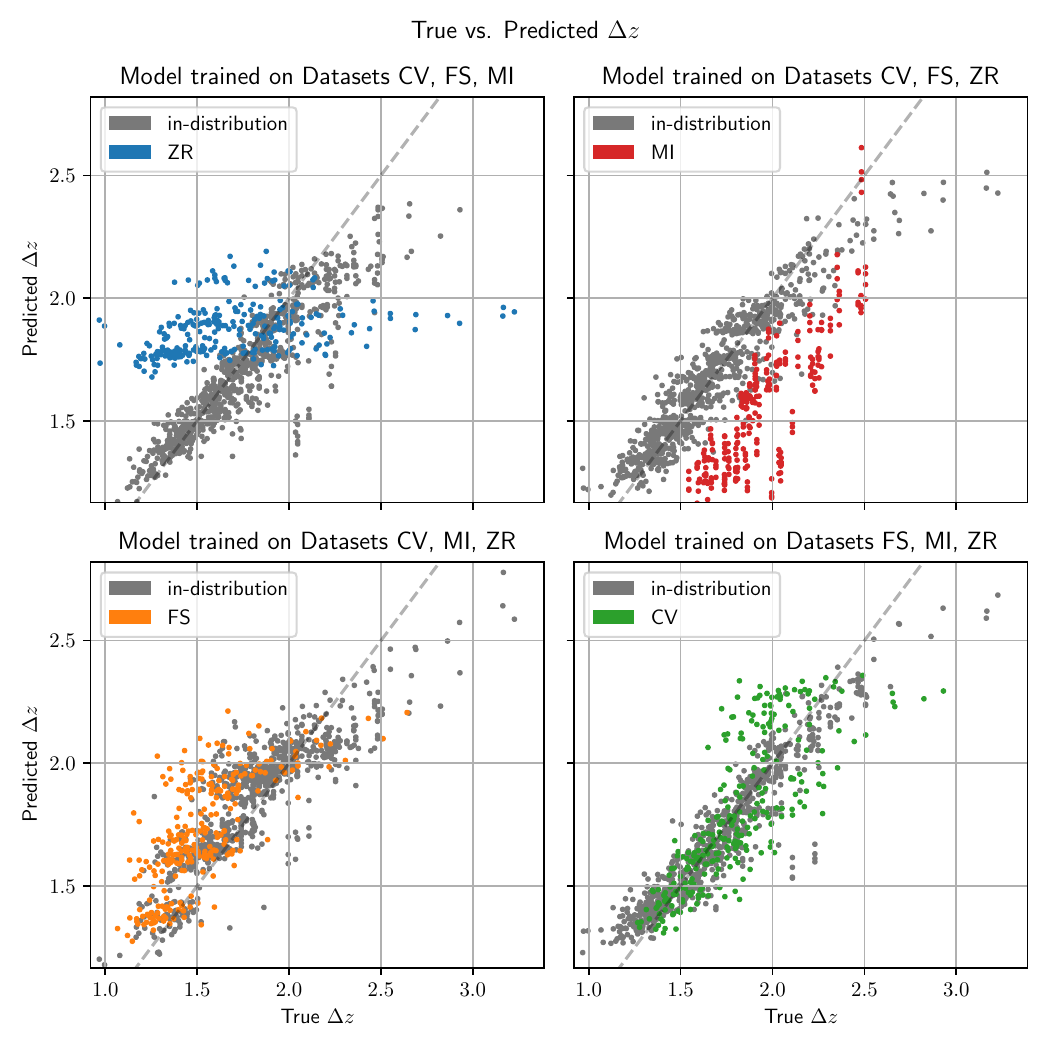}
    \caption{Predicted vs. true $\Delta z$ for models trained on three datasets. Note that unlike the training data, which was split equally between the three simulators used during training, the test datasets plotted here remain at a fixed size, in order to better compare the test performance across models.} 
\label{fig:triple_dataset_ood_scatter}
\end{figure*}

The predicted vs. true value of $\Delta z$ for models trained on two and three datasets is plotted in Figures \ref{fig:double_dataset_ood_scatter} and \ref{fig:triple_dataset_ood_scatter}, respectively. Accuracy for in-distribution datasets (shown in gray) remains high, with an average MSE of $0.02$ and $0.03$ for the double and triple dataset models, respectively. Similarly, parameter recovery for out-of-distribution data (shown in color) remains poor, with an average MSE of $0.17$ and $0.13$ across all out-of-distribution dataset / model pairs for the double and triple dataset models, respectively. Compared to the average out-of-distribution MSE of the single dataset models $0.32$, adding data from additional simulators to the suite of training data results in a clear improvement in an AI model's ability to generalize. In addition to being quantitatively better as a function of the number of included simulators, the out-of-distribution performance is visually improved between Figures \ref{fig:single_dataset_ood_scatter}, \ref{fig:double_dataset_ood_scatter}, and \ref{fig:triple_dataset_ood_scatter}. While parameter recovery never quite reaches the level of accuracy achieved for in-distribution data, a reduction in MSE of $\approx60\%$ from the single-dataset case to the triple-dataset case is a striking improvement. 

\begin{figure}
\includegraphics[width=\columnwidth]{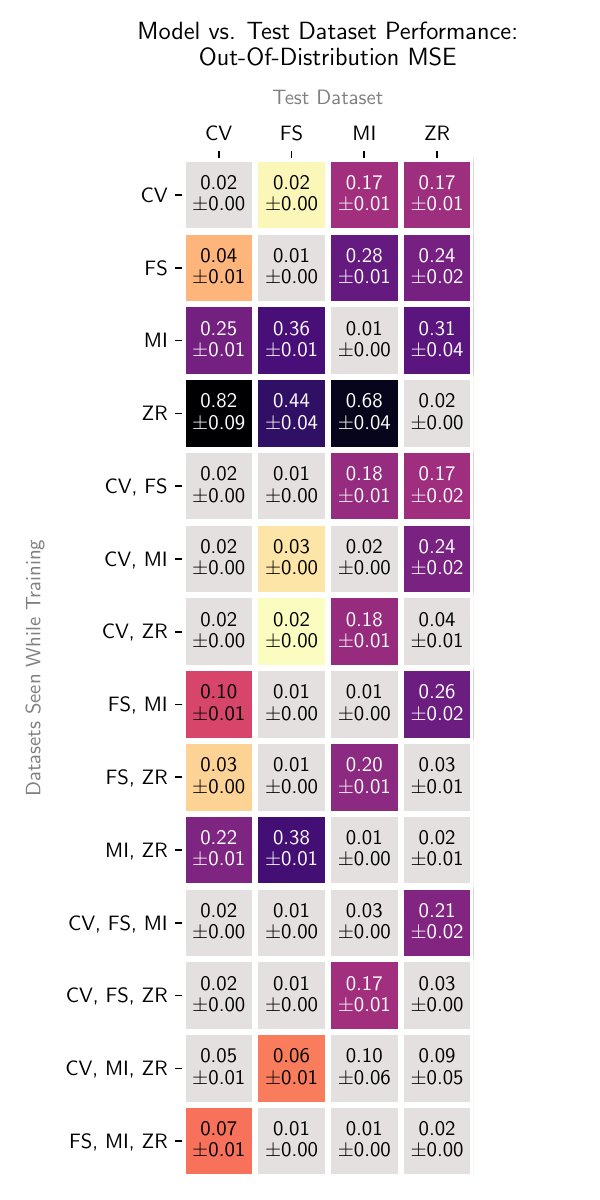}
    \caption{Average prediction MSE per test dataset of CNNs trained on different combinations of training sets, $\pm$ standard error over 8 folds. In this figure, the row represents the simulators represented in the model's training data and the column represents the test dataset. In-distribution model-dataset pairs are shown in gray and out-of-distribution pairs are shown in color. Error values $<0.005$ have been rounded to $0.00$.} 
    \label{fig:multi_dataset_ood}
\end{figure} 

Figure \ref{fig:multi_dataset_ood} plots shows the average MSE prediction error for out-of-distribution datasets across all models, $\pm$ the standard error across folds. It is worth noting the impact of including Dataset \zr, the most dissimilar from the other datasets and the dataset generated from the most approximate semi-numerical simulator, affects the out-of-distribution performance of the other three datasets. The results here are inconclusive: in some cases including Dataset \zr\ improves performance (the MSE when predicting on Dataset \miz\ is $0.28\pm0.01$ for the model trained on \fsp,  vs. $0.20\pm0.01$ for the model trained on both \fsp\ and \zr), but in other cases it does not significantly impact performance at all (the MSE when predicting on Dataset \fsp\ is $0.02\pm0.00$ for both the model trained on \cv\ and the model trained on both \cv\ and \zr). In only one case does the addition of \zr\ actively hurt performance: the MSE when predicting on Dataset \fsp\ is $0.03\pm0.00$ for the model trained on \cv\ and \miz,  vs. $0.06\pm0.01$ for the model trained on \cv, \miz, and \zr. But to truly compare how adding data from a new distribution impacts out-of-distribution performance, we must consider how adding an additional dataset on average impacts the out-of-distribution performance.

\begin{figure}
    \includegraphics[width=\columnwidth]{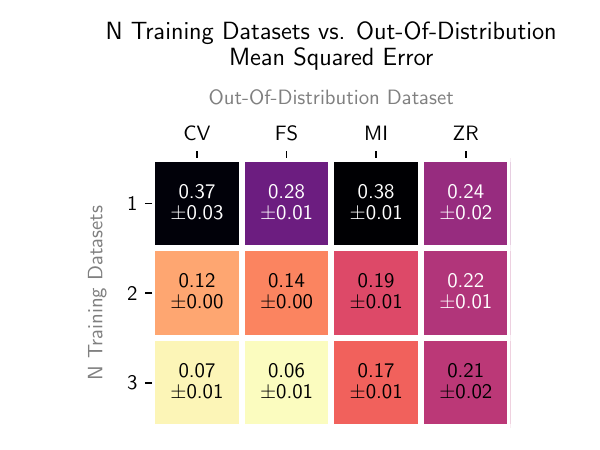}
    \caption{Out-of-distribution performance heatmap averaged across number of simulators represented in the training data, $\pm$ standard error over 8 folds. This figure groups the data presented in Figure \ref{fig:multi_dataset_ood} by number of simulators represented in the training pool then averages along the columns. Note that the averaging takes place over a very small sample size, as the number of out-of-distribution prediction instances are limited by the fact that we only have four available datasets. Error values $<0.005$ have been rounded to $0.00$.}
    \label{fig:avg_per_n_datasets_ood}
\end{figure}

To compare the average out-of-distribution MSE for models trained on one, two, and three datasets, we group the columns of Figure \ref{fig:multi_dataset_ood} by the number of datasets included in the training data then average across the out-of-distribution values to obtain Figure \ref{fig:avg_per_n_datasets_ood}. For example, the value of cell (1, \cv) in Figure \ref{fig:avg_per_n_datasets_ood} is the average of cells ([\fsp], \cv), ([\miz], \cv), and ([\zr], \cv) in Figure \ref{fig:multi_dataset_ood}; similarly, cell (3, \cv) in Figure \ref{fig:avg_per_n_datasets_ood} is simply cell ([\fsp, \miz, \zr], \cv) in Figure \ref{fig:multi_dataset_ood}. Therefore, Figure \ref{fig:avg_per_n_datasets_ood} shows us the average out-of-distribution prediction error as a function of the number of distributions included in the data. 
Figure \ref{fig:avg_per_n_datasets_ood} shows that for test datasets \cv, \fsp, and \miz, the average prediction MSE of unseen datasets is significantly improved when an additional distribution is added to the pool of training data. For Dataset \zr, the results are more inconclusive, with the improvements in performance falling outside the range of significance. It also does not, however, appear to significantly worsen performance, which is also important when considering the impact of training models meant for real data on simulations. It is notable that \zr\ is the most dissimilar of the four simulators, as well as the least physically motivated; this may explain the lack of improvement in its out-of-distribution performance in relation to the other simulator datasets.
Overall, these findings imply that increasing the diversity of the training data by including data drawn from multiple different sources will be important when using AI models to analyze real-world data.

\section{Discussion} 
\label{sec:discussion}

\subsection{Cross-Simulator Performance Transfer}
\label{sec:x_sim_performance_transfer}

\begin{figure}
    \includegraphics[width=\columnwidth]{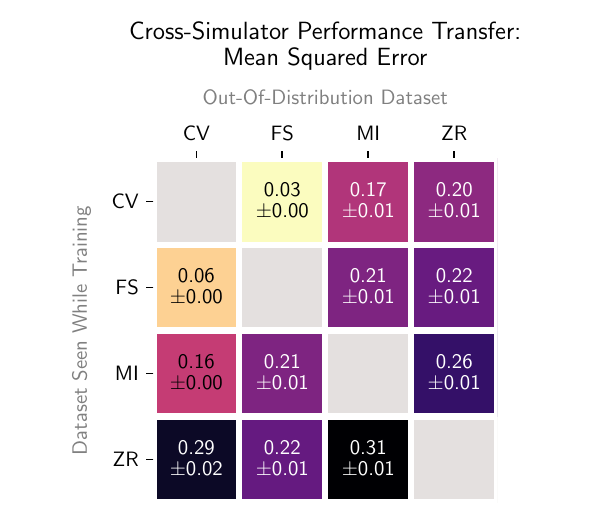}
    \caption{Cross-simulator performance transfer. This plots the average MSE performance for models where the row dataset is included in the training data while the column dataset is not, $\pm$ standard error over 8 folds. For example, cell (\cv, \fsp) is the performance on Dataset \fsp\ averaged across all models trained on \cv\ but not \fsp\ (aka, models trained on \cv; \cv\ and \miz; \cv\ and \zr; and \cv, \miz, and \zr). This metric does not disentangle the effects of multiple distributions being present in the training data, but rather gives a general sense of how transferable the model's knowledge is between two simulators. Error values $<0.005$ have been rounded to $0$.} 
    \label{fig:cross_sim_performance}
\end{figure}

Figure \ref{fig:cross_sim_performance} estimates how well a model trained on data from one simulator will generalize to data from another unseen simulator by averaging the out-of-distribution prediction MSE by included dataset. In other words, Figure \ref{fig:cross_sim_performance} was derived in the same way as Figure \ref{fig:avg_per_n_datasets_ood}, except in \ref{fig:cross_sim_performance} the columns were averaged by whether a given dataset was seen during training or not. For example, the value of cell (\fsp, \cv) in Figure \ref{fig:cross_sim_performance} is the average of cells ([\fsp], \cv), ([\fsp, \miz], \cv), ([\fsp, \zr], \cv), and ([\fsp, \miz, \zr], \cv), in Figure \ref{fig:multi_dataset_ood}. Figure \ref{fig:cross_sim_performance} does not uncouple the contributions of individual simulators to the model's ability to generalize to data from the unseen simulator; instead, it provides a rough metric of how well the model is able to apply knowledge of one dataset to another in order to perform accurate parameter recovery. 

Figure \ref{fig:cross_sim_performance} shows that including certain datasets in the training pool allows the model to infer the parameters of some unseen datasets better than others. For example, models trained on Dataset \cv\ tend to perform better on an out-of-distribution Dataset \fsp\ (with an average MSE of $0.03\pm0.00$) than on out-of-distribution Datasets \miz\ or \zr\ (with an average MSE of $0.17\pm0.01$ and $0.20\pm0.01$, respectively). It is unclear why exactly the model is able to generalize better between Datasets \cv\ and \fsp\ and others, beyond suggesting that the model places less importance on the distribution of bubble sizes in comparison to other features of the datasets, as the only difference between Datasets \cv\ and \fsp\ is in how ionized bubbles are flagged.

Exactly quantifying this cross-simulator performance transfer in a way that would allow one to predict how two simulators might generalize to one another in advance is difficult, and outside the scope of this work. 

\subsection{Quantity vs. Quality}
\label{sec:quantity_quality}

A natural question these results inspire is one of data quantity vs. quality. Figure \ref{fig:cross_sim_performance} tells us that certain datasets are more valuable than others when predicting on data from an unseen distribution. For example, if we take Dataset \cv\ to be our proxy for the real universe, a model trained on Dataset \fsp\ would give significantly better parameter estimates than one trained on Dataset \zr. 

However, Figure \ref{fig:avg_per_n_datasets_ood} tells us that the diversity of the training dataset is also important. When recovering EoR parameters, we cannot be sure which of our simulators will allow our model to best generalize to real data. We may have a fuzzy spectrum of most to least physically motivated (for example, \cmfast\ is more physically motivated than \zreion, and therefore we assume it to be more reflective of the real universe), but for neighbors on that spectrum, which simulator is best suited to allow a model to accurately generalize to 21cm observations is an open question (for example, it is unclear whether central-voxel or full-sphere flagging is better reflective of reionization). Therefore, we can only rely on the out-of-distribution performance averaged as a function of the number of distributions (aka, simulators) present in the training data. With this in mind, Figure \ref{fig:avg_per_n_datasets_ood} tells us that, on average, out-of-distribution performance improves with increased training data diversity, thus suggesting it is best to include data from as many sources as possible.

More approximate simulators can be run more quickly to produce greater quantities of data. The diversity argument would favor including this data in our training set, while the quality argument states there's a limit to how useful this inclusion can be corresponding to how closely this simulator overlaps with the true distribution of possible universes. Figure \ref{fig:avg_per_n_datasets_ood} seems to imply there is no negative impact, and some potential benefit, to including high volumes of lower-accuracy data, such as that generated by \zreion, in the training pool. 

There is, as always, a balance between quantity and quality to be found. As an extreme example, including ``lightcones'' composed of random Gaussian noise in the training pool will not contribute any meaningful physical information to the model, and therefore will not improve parameter recovery, despite this technically increasing the diversity of the training pool. This raises an important question: which datasets are ``good enough'' to be worth including in the training pool? 

Because we do not have access to complete information about the ground truth (aka, the real universe), this is a difficult question to answer. Future work may address this question by studying how accurately models trained on semi-numeric simulators can recover the parameters of data generated from high-fidelity numeric simulators.

Still, with the assumption that our simulators at least capture some useful physical information, our results support the creation of large, multifaceted training datasets made up of data drawn from simulators with a range of numerical complexities over the curation of small, highly accurate datasets sourced from one or two models of reionization. This has implications for future works involving AI-based EoR parameter inference, as well as ongoing data analysis efforts that involve the use of AI, such as LOFAR's introduction of ML-GPR into their analysis pipeline (\citealt{Mertens2023}, \citealt{Mertens2025}, \citealt{Munshi2025}).

\subsection{Future Work} 
\label{sec:future_work}

An obvious continuation of this work would be to produce additional datasets from new EoR simulation algorithms to see whether the trend of training set diversity improving out-of-distribution performance in accordance with the trend shown in Figure \ref{fig:avg_per_n_datasets_ood} continues. This would be challenging due to the time and resources involved in data generation, as well as due to the lack of software maintenance many older reionization simulation packages suffer from. Generating a small set of reionization histories from highly accurate radiative transfer simulators would be especially useful in characterizing how AI models trained on more approximate semi-numerical simulators respond to more realistic data.

Another route would be to perform the same experiment with modeled foregrounds and noise. It is possible, though somewhat unlikely, that the removal of foreground-contaminated k-modes from the data would cause our AI models to focus less on simulator-specific features, leading to greater out-of-distribution performance. This would, of course, come with some corresponding decrease in overall performance as information is lost from the data. How noise and foregrounds impact model generalization likely hinges on how much simulator-specific information is contained in k-modes we expect to be contaminated by foregrounds.

Another followup involves precisely defining and quantifying some metric of ``simulator similarity'' that would allow one to accurately predict how training a model on data from one simulator will affect performance on data from an unseen second simulator. Figure \ref{fig:cross_sim_performance} gives a rough metric of which datasets the model sees as comparable, but does nothing to predict how training an AI on any given distribution would impact its performance on a second unseen distribution. Studying the topological space of these simulators and how they overlap (or fail to overlap) may yield useful information in this vein.

In addition to further characterizing our simulators, we can also characterize the models themselves, identifying what features of the data factor into their predictions and why. This falls under the umbrella of explainable AI (XAI) research: work to develop new methods and algorithms capable of shedding light on the ``black box'' of a machine learning model. Gradient-based techniques such as Saliency Mapping and Integrated Gradients have been used in prior research involving AI-based EoR parameter recovery in an attempt to identify which parts of the input images feature most heavily in the model's decision making (\citealt{Lahiry2025}, \citealt{LaPlante2019}, \citealt{Prelogovic2021}). We have found these localized feature attribution methods to be insufficient for this task: there are visible differences in the interpretability metrics between models trained with more simulators when compared with model trained ony a single simulator, but they vary significantly from sample to sample and no clear trend presents itself (see Appendix \ref{appendix:interp}). Future research developing non-localized interpretability methods is already in the works.

The ultimate goal of this simulation-based AI parameter inference of the EoR would be to develop a suite of training data that maximizes an AI's ability to perform well on unseen reionization scenarios. This would allow for the eventual use of AI on analysis of real-world EoR data. 

\section{Conclusions}
AI is a promising tool for constraining parameters of reionization, but it is not without its pitfalls. AI models are known to struggle with out-of-distribution prediction tasks, which is precisely the type of task analyzing real-world 21cm data will require.

Regressing on parameters of the EoR, both via AI or more traditional Bayesian methods, has largely been studied only in the context of one parameter space. This is a problem because different models of reionization have different strengths and weaknesses. A method that combines the strengths of multiple distributions while compensating for their weaknesses would allow for a more unbiased, simulator-independent treatment of the data.

This work investigates the cross-simulator performance of AI models, and finds that in general AI models trained on data from a single simulator of reionization fail to generalize to data from a second unseen simulator. When the training data is diversified by the inclusion of data from multiple simulators in the training pool, the AI model's performance on a held-out dataset improves. This implies that a diverse training set with data drawn from many different EoR simulators is essential to accurately predicting on out-of-distribution data, such as actual 21cm observation data. 

\section*{Acknowledgments}

The authors acknowledge support from U.S National Science Foundation awards \#2106510 and \#2509340.  This research was funded in part by the NASA EPSCoR Research Infrastructure Development program, award 80NSSC22M0040.  JS also received support from the NASA Rhode Island Space Grant Program, award 80NSSC20M0053.  This research was conducted using computational
resources and services at the Center for Computation and Visualization, Brown University.  The authors also thank the anonymous referee for helpful suggestions that improved the quality of this manuscript.  

Disclosure: Stephen Bach is an advisor to Snorkel AI, a company that provides software and services for data-centric artificial intelligence.

\section*{Data Availability}
The data underlying this work are available on request to the authors.

\bibliographystyle{mnras}
\bibliography{bib}

\appendix
\section{Detailed Dataset Generation Parameters}\label{appendix:dataset_params}

Table \ref{tab:compare_datasets_detailed} details the parameter ranges used when generating our datasets.

\begin{table*}
\label{tab:compare_datasets_detailed}
    \centering

    \begin{tabular}{lcccccc}
        \hline \hline 
        \multicolumn{7}{c}{\textbf{Dataset Generation Parameters}}\\
        \hline \hline
        \multicolumn{2}{c}{Parameter} & Unit & Value & Value & Value & Value\\
         &  &  & (Dataset \cv) & (Dataset \fsp) & (Dataset \miz) & (Dataset \zr)\\
        \hline
        \multicolumn{2}{c}{HII\_DIM} & - & 256 & 256 & 512 & 256 \\
        \multicolumn{2}{c}{BOX\_LEN} & Mpc & 1000 & 1000 & 2000 & 1000 \\
        \multicolumn{2}{c}{USE\_MASS\_DEPENDENT\_ZETA} & - & True & True & False & -\\
        \multicolumn{2}{c}{IONISE\_ENTIRE\_SPHERE} & - & False & True & False & -\\
        \hline \hline
        Parameter & Flat Prior & Unit & Allowed Range & Allowed Range & Allowed Range & Allowed Range\\
         &  &  & (Dataset \cv) & (Dataset \fsp) & (Dataset \miz) & (Dataset \zr)\\
        \hline
        $f_{*,10}$ & log & - & $[0.001, 1]$ & $[0.01, 1]$ & - & -\\
        $\alpha_{*}$ & linear & - & $[-0.5, 1]$ & $[-0.5, 1]$ & - & - \\
        $f_{\text{esc},10}$ & log & - & $[0.001, 1]$ & $[0.01, 1]$ & - & -\\
        $\alpha_{\text{esc}}$ & linear & - & $[-1, 0.5]$ & $[-1, 0.5]$ & - & -\\
        $M_{\text{turn}}$ & log & M$_\odot$ & $[10^{8}, 10^{10}]$ & $[10^{8}, 10^{10}]$ & - & -\\
        $t_{*}$ & linear & - & $[0, 1]$ & $[0, 1]$ & - & -\\
        $\zeta$ & linear & - & - & - & $[10, 250]$ & -\\
        $T^\text{min}_\text{vir}$ & log & K & - & - & $[10^4, 10^6]$ & -\\
        $\bar{z}$& linear & - & - & - & - & $[7,13]$ \\
        $\alpha$ & linear & - & - & - & - & $[-0.5,0.5]$\\
        $k_0$ & linear & $h\text{Mpc}^{-1}$ & - & - & - & $[0.1,1]$\\
        \hline \hline
    \end{tabular}
    
    \caption{Settings and parameter ranges used to generate our datasets. If not otherwise specified, the default values were used. The allowed ranges were informed by \protect\cite{Park2019} and \protect\cite{Battaglia2013}, but were adjusted to produce datasets with overlapping $\Delta z$ distributions (see Figure \ref{fig:dur_distribution}).}
\end{table*}

\section{Detailed Model Architecture}\label{appendix:model_architecture}

Table \ref{tab:model_architecture} describes the detailed architecture of our model. Layer names and parameters are those used in the PyTorch implementation of our model.

\begin{table*}
    \label{tab:model_architecture}
    \centering
    \begin{tabular}{cccc}
        \hline \hline
        \multicolumn{4}{c}{\textbf{Model Architecture}} \\
        \hline \hline
        \textbf{Name} & \textbf{Layer Class} & \textbf{Input Params} & \textbf{Trainable Params}\\
        \hline
        \multirow{4}{60pt}{\texttt{conv\_stack\_1}} & Conv2d & \texttt{out\_channels=16} & 4336 \\
        & ReLU & - & - \\
        & BatchNorm2d & - & 32 \\
        & MaxPool2d & \texttt{kernel\_size=2} & -\\
        \hline
        \multirow{4}{60pt}{\texttt{conv\_stack\_2}} & Conv2d & \texttt{out\_channels=32} & 4640 \\
        & ReLU & - & - \\
        & BatchNorm2d & - & 64 \\
        & MaxPool2d & \texttt{kernel\_size=2} & - \\
        \hline
        \multirow{4}{60pt}{\texttt{conv\_stack\_3}}& Conv2d & \texttt{out\_channels=64} & 18496\\
        & ReLU & - & - \\
        & BatchNorm2d & - & 128\\
        & MaxPool2d & \texttt{kernel\_size=2} & - \\
        \hline
        \multirow{2}{60pt}{\texttt{global\_maxpool}} & MaxPool2d & \texttt{kernel\_size=32} & - \\
        & Flatten & - & - \\
        \hline 
        \multirow{3}{60pt}{\texttt{linear\_stack\_1}} & Dropout & - & - \\
        & Linear & \texttt{out\_features=200} & 13000\\
        & ReLU & - & - \\
        \hline
        \multirow{3}{60pt}{\texttt{linear\_stack\_2}} & Dropout & - & - \\
        & Linear & \texttt{out\_features=100} & 20100\\
        & ReLU & - & - \\
        \hline
        \multirow{3}{60pt}{\texttt{linear\_stack\_3}} & Dropout & - & - \\
        & Linear & \texttt{out\_features=20} & 2020 \\
        & ReLU & - & - \\
        \hline
        \texttt{output} & Linear &  \texttt{out\_features=1} & 21 \\
        \hline \hline
        \multicolumn{3}{r}{Total no. trainable parameters:} & 62837 \\
        \hline \hline
    \end{tabular}
    \caption{A detailed description of our CNN architecture, as implemented in PyTorch. All Conv2d layers use the parameters \texttt{kernel\_size=3} and \texttt{padding=same}. All Dropout layers use \texttt{p=0.2}.} 
\end{table*}

\section{Discussion of Interpretability Methods}
\label{appendix:interp}
In our exploration of explainable AI (XAI) techniques such as Saliency Maps and Integrated Gradients for our data and models, we've found none of the standard tools useful for the task of analyzing EoR brightness temperature maps.  The major limitation is largely that these tools highlight \emph{localized} features of an image that most affect the predictions, whereas cosmological analyses are generally driven by image-wide statistical features (hence, for example, the value of the cosmological power spectrum). We now have a grant to research new non-local interpretability methods suitable for the task at hand (\url{https://www.nsf.gov/awardsearch/show-award?AWD\_ID=2509340}) which will follow up on this work.

\begin{figure*}
    \centering
    \includegraphics[height=0.4\textheight]{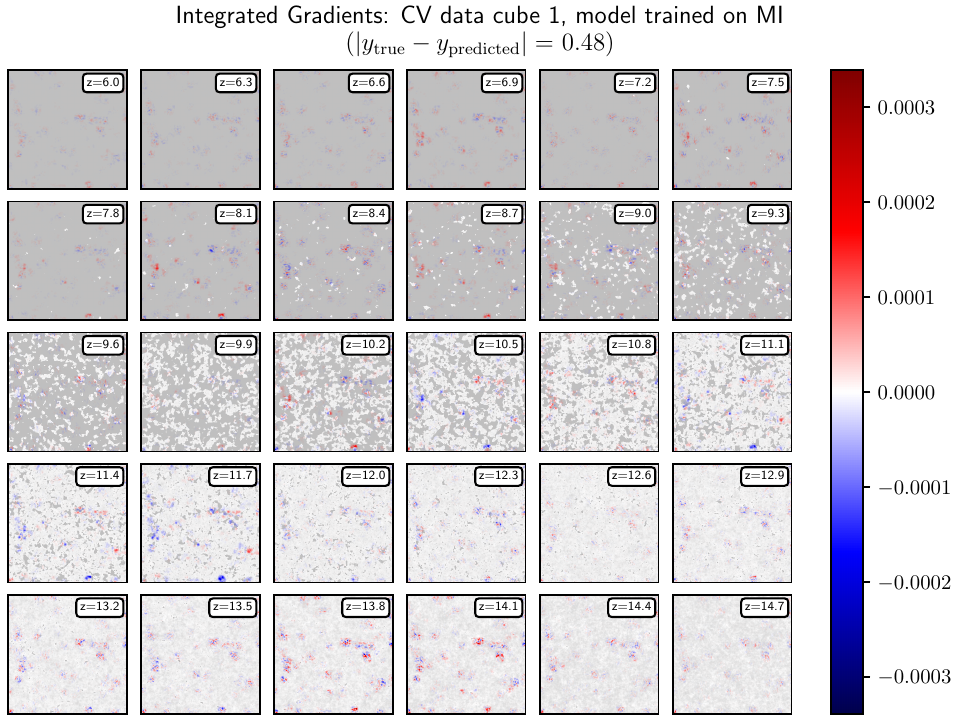}
    \includegraphics[height=0.4\textheight]{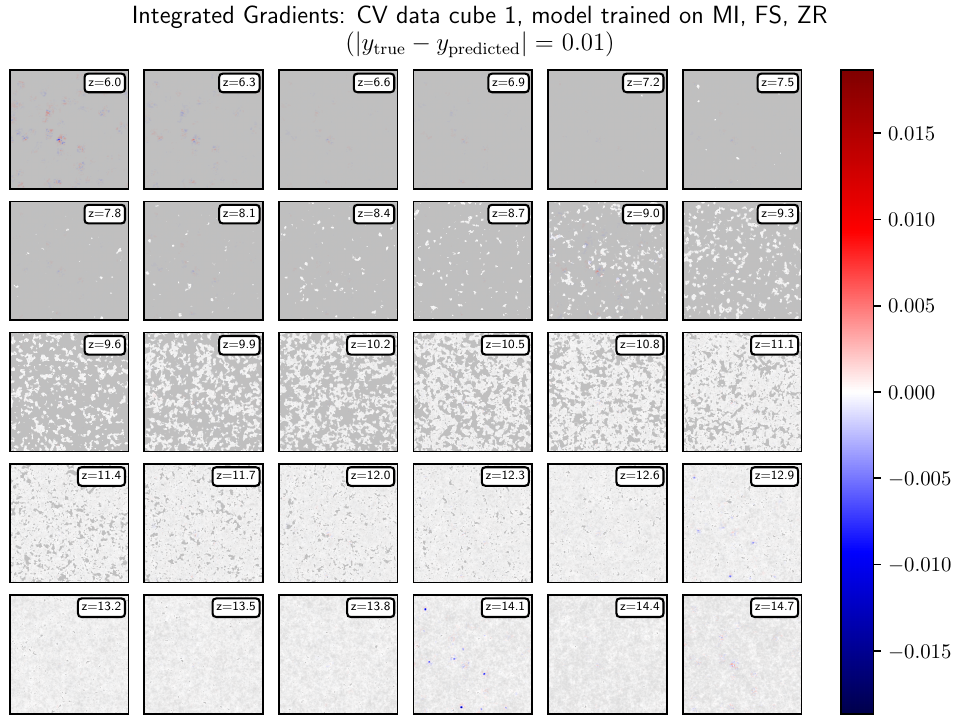}
    \caption{Integrated Gradient map for a single and triple simulator model predicting on an out-of-distribution sample from Dataset \cv. For ease of viewing, a gaussian smoothing filter was applied to the attribution map with $\sigma=1$.}
    \label{fig:ig_cv_1}
\end{figure*}

\begin{figure*}
    \centering
    \includegraphics[height=0.4\textheight]{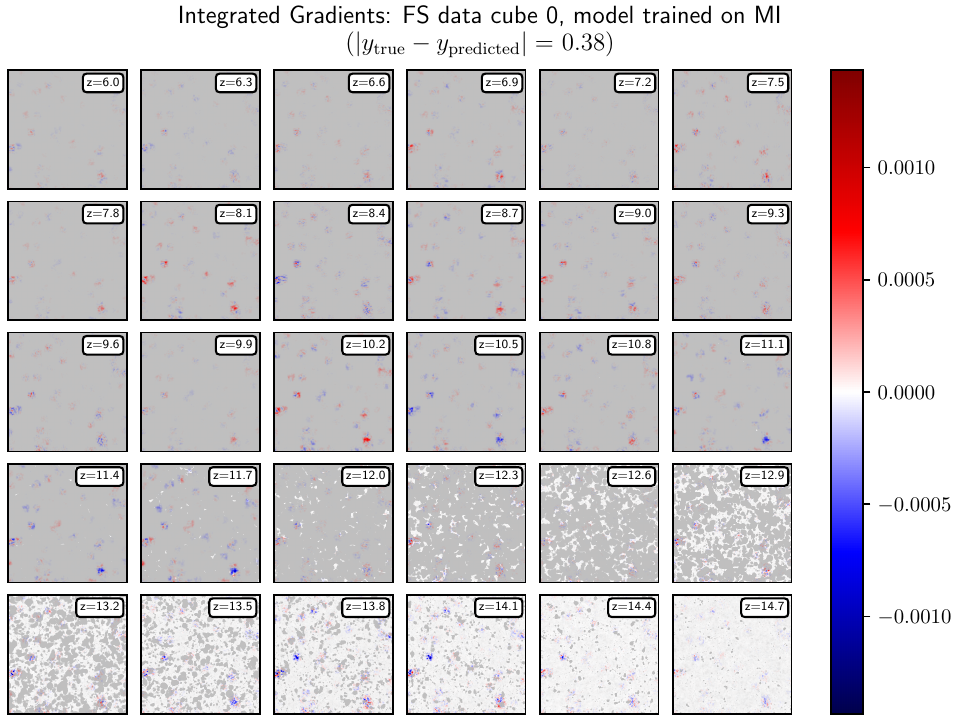}
    \includegraphics[height=0.4\textheight]{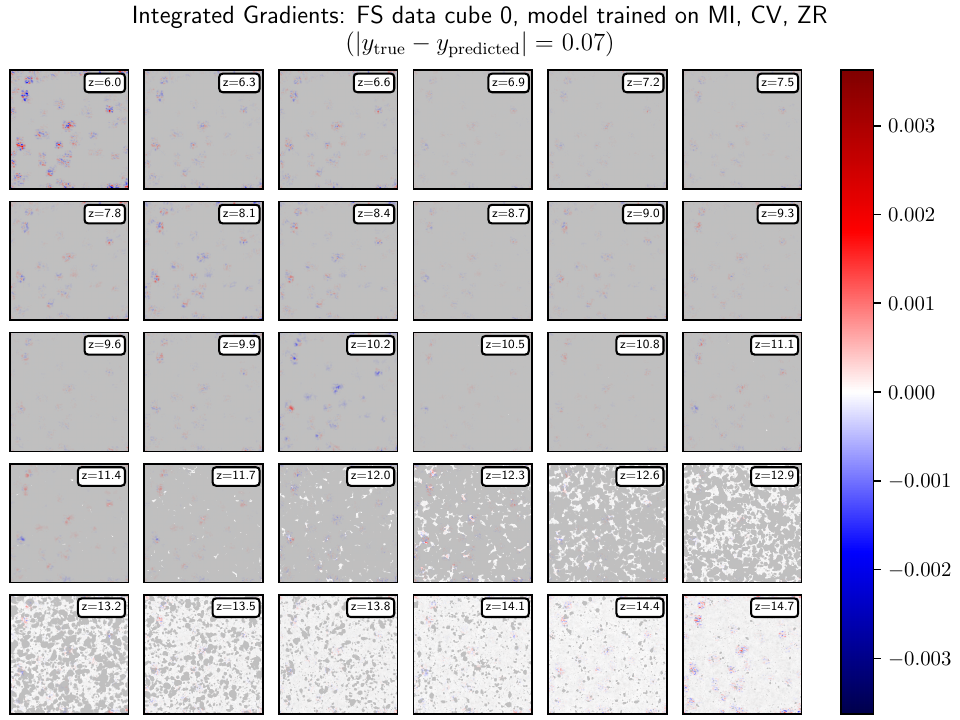}
    \caption{Integrated Gradient map for a single and triple simulator model predicting on an out-of-distribution sample from Dataset \fsp, as in Figure \ref{fig:ig_cv_1}.}
    \label{fig:ig_fs_0}
\end{figure*}

\begin{figure*}
    \centering
    \includegraphics[height=0.4\textheight]{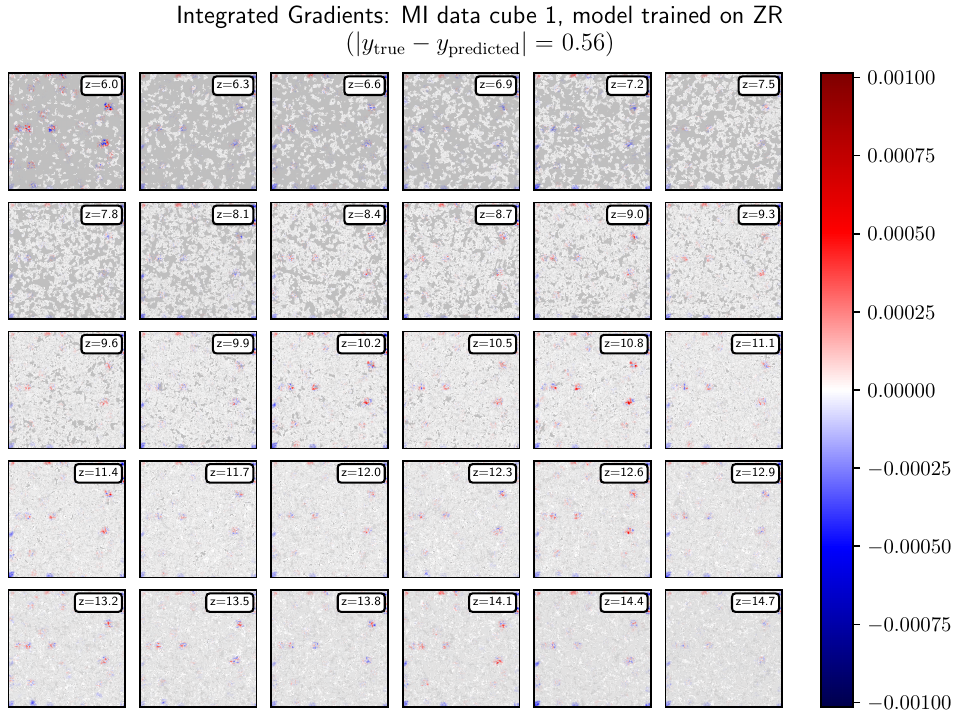}
    \includegraphics[height=0.4\textheight]{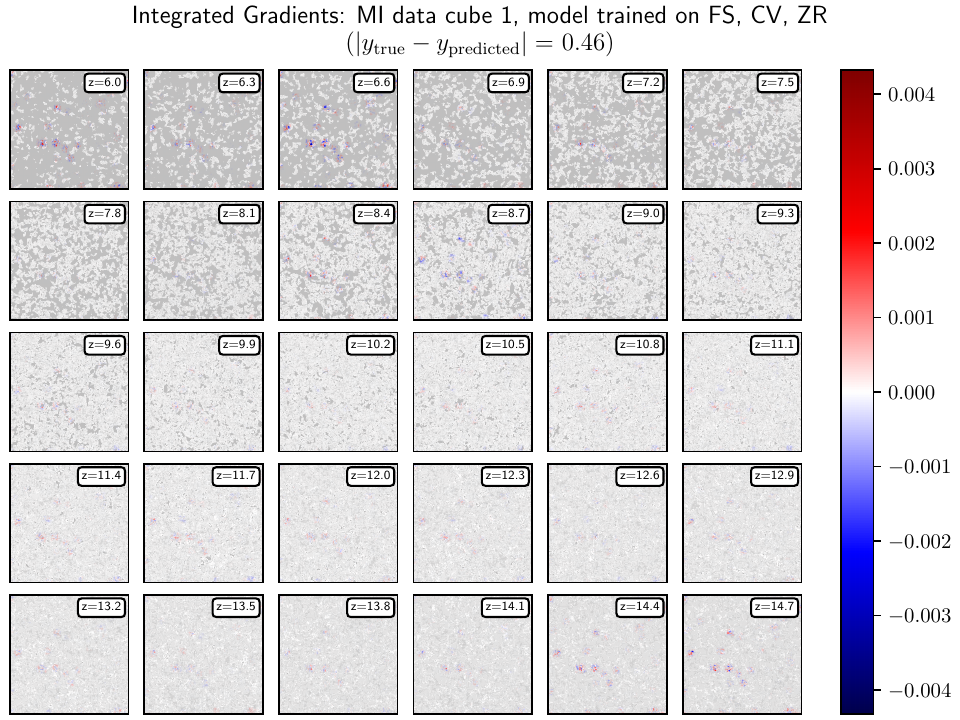}
    \caption{Integrated Gradient map for a single and triple simulator model predicting on an out-of-distribution sample from Dataset \miz, as in Figure \ref{fig:ig_cv_1}.}
    \label{fig:ig_mi_1}
\end{figure*}

As a demonstration of the difficulty in interpreting these gradient-based XAI techniques, we include here Figures \ref{fig:ig_cv_1}, \ref{fig:ig_fs_0}, and \ref{fig:ig_mi_1}, each of which shows the Integrated Gradients attribution maps for a given out-of-distribution sample passed through two models: a single-simulator model and a triple-simulator model. We specifically chose samples where the performance improved from the single to the triple simulator model, to see if differences in the attribute mapping could reveal why the out-of-distribution performance improved. It is difficult to glean meaningful insights from these maps; the integrated gradients look different between the two models, but there is no obvious trend that can be invoked to explain the improvements achieved by the triple-simulator models.  Similar results are seen with Saliency Maps.

XAI methods such as Saliency Mapping and Integrated Gradients also suffer from a more fundamental problem, which is that they can only be applied to individual inputs, and do not, in and of themselves, make any predictions about what the model will find relevant about an unseen sample. That is, they apply a post-hoc explanation to individual samples while not making any statistical inferences about the model itself. One would need to create a new dataset of attribution maps and perform a statistical analysis of that to draw any conclusions about the model itself, a project that is beyond the scope of this work.

Ultimately, the inconclusiveness of these results and our new research thrust on non-local interpretability lead us to postpone a discussion of interpretability to future work.

\bsp
\label{lastpage}
\end{document}